\documentclass[referee]{aa}
\usepackage{graphicx}
\begin{document}

   \thesaurus{9 (02.13.2; 02.23.1; 06.01.2; 06.03.2; 06.15.1; 06.21.1)}

   \title{Wide-Spectrum Slow Magnetoacoustic Waves in Coronal Loops}

   \author{D. Tsiklauri and V.M. Nakariakov}

   \offprints{D. Tsiklauri (tsikd@astro.warwick.ac.uk)}

   \institute{Physics Department,
   University of Warwick, Coventry, CV4 7AL, England}

   \date{\today}
\titlerunning{Wide-spectrum slow waves in loops}
\authorrunning{D.~Tsiklauri \& V.M.~Nakariakov}
   \maketitle

   \begin{abstract}
A model interpreting variations of EUV
brightness upward propagating  in solar coronal loops as slow magnetoacoustic
waves is developed. A loop is considered to have a non-zero plane
inclination angle and offset of circular loop centre from the baseline.
The model also incorporates effects of dissipation and gravitational
stratification. A linear evolutionary equation is derived and applied to
investigation of slow wave dynamics. Both the non-zero plane
inclination and the offset decrease the growth of the wave amplitude 
due to stratification.
It is shown that wide spectrum slow
magnetoacoustic waves, consistent with currently available observations
in the low frequency part of the spectrum, can provide the
rate of heat deposition sufficient to heat the loop. In this
scenario, the heat would be deposited near the loop footpoints
which agrees with the current observational data.
\end{abstract}

\section{Introduction}

Identification of possible mechanisms of heating of solar coronal
plasma and acceleration of solar wind is among the unsolved problems
of solar physics. In the last two decades there has been intensive
theoretical work undertaken to advocate the idea of coronal plasma
heating by magnetohydrodynamic (MHD) waves. In particular, it has
been suggested that coronal loops could be heated by dissipation
of high frequency slow magnetoacoustic waves (Porter, Klimchuk \&
Sturrock 1994a,b; Laing \& Edwin 1995). However, 
up until recent observational advances,
these studies have been poorly motivated
because of 
the absence of observational evidence of the MHD wave activity in the
solar corona.

Recent observations performed with SOHO/EIT and TRACE imaging
telescopes have established:

1. Existence of upward propagating perturbations of EUV brightness
in long coronal loops at the bandpasses EIT 195\AA\ (Berghmans \&
Clette 1999) and TRACE 171\AA\ (Berghmans et al. 1999, De Moortel,
Ireland \& Walsh  2000), and EIT 195\AA\ and TRACE 171\AA\
simultaneously (Robbrecht et al. 2001). Probably, the same
phenomenon was observed by Nightingale, Aschwanden \& Hurlburt
(1999) in TRACE 171\AA\ .

2. The projected velocities of these plasma density perturbations
are about 65-165 km/s, which is below the sound speed
corresponding to the temperature of the loop.

3. The amplitude is usually 1-3\% of the background density,
reaching 5-6\% in some observations.

4. The propagating disturbances are quasi-periodic with the
periods of 3-20 min. There is no a well-defined correlation
between the periods and physical parameters of the loops detected.

5. The disturbances are detected in the loop legs near footpoints
and always propagate upwards. There are no observations of waves
propagating downwards.

Two rival interpretations of the propagating disturbances have
been suggested. In the first (Reale, Peres \& Serio 2000), the
disturbances are considered as field-aligned flows of matter,
generated by the siphon effect. In the second (Nakariakov et al.
2000), the propagating disturbances are associated with slow
magnetoacoustic waves.
The wave-based theory is similar to the model developed for the
interpretation of compressive perturbations observed in polar
plumes (Ofman, Nakariakov \& Deforest 1999, Ofman, Nakariakov \&
Sehgal 2000) in terms of slow magnetoacoustic waves.
 Both theories are consistent with the
observations and with the MHD theory, but the flow-based model has
difficulties with explanation of the periodicity and the symmetry
of the front and back edges of the disturbance. Also, the
observations do not show a change (acceleration) in the
propagating disturbance speed, predicted by the flow-based theory.
In contrast, the wave-based theory explains all the observational
findings and, in particular, justifies the absence of downward
propagating disturbances.

In the wave-based theory (Nakariakov at al. 2000), it has been
established that the main physical factors, which are important in the
slow magnetoacoustic wave dynamics in long coronal loops, include:
(i) gravitational stratification, which is responsible for wave
amplitude amplification; (ii) dissipation, which leads to the
conversion of wave kinetic energy into heat in the large wave
numbers part of the spectrum, and (iii) nonlinearity, which
generally yields wave front steepening by generation of higher
harmonics, leading to enhanced dissipation. An evolutionary
equation for the slow magnetoacoustic wave amplitude has been
derived, which includes these three effects. However, in order to
catch the physical essence of the problem without further
complications, simple geometry of the loop, namely, zero
inclination angle and zero offset of the centre from the solar
surface has been assumed.

In contrast, observations show that the plane of a loop is not
necessary perpendicular to the Sun's surface (a non-zero
inclination) and the centre of the loop circle is usually situated
below the surface (a non-zero offset) (e.g. Aschwanden et al. 1999).
Consequently, a more elaborate theory of the slow magnetoacoustic
waves in coronal loops, which includes the effects of the non-zero
inclination and the non-zero offset, is required.

Soft X-ray and EUV coronal telescopes provide the possibility to
determine the distribution of density, temperature and heat
deposition along a loop. Priest et al. (2000), using Yohkoh/SXT
observations concluded that hot, soft X-ray loops were heated
uniformly. Aschwanden et al. (1999), Aschwanden, Nightingale \& Alexander 
(2000) and  Aschwanden, Schrijver \& 
Alexander (2001) analyzing SOHO/EIT and
TRACE EUV observations argued that cooler loops were heated
non-uniformly, near the loop footpoints. The specific mechanism for
the energy deposition remains unknown and could, in principle,
be associated with dissipation of slow magnetoacoustic waves.  

De Moortel et al. (2000) have estimated the energy of slow waves
observed to be insufficient to heat a loop. This result was
based upon the assumption of the wave to be a single harmonic with the
period of several minutes. But, currently available resolution of
the telescopes does not allow us to detect periodic motions with
the periods shorter than a few minutes, even if they are presented
in the spectrum. On the other hand, the absence of well-defined
period in the loop waves suggests that their spectrum is
probably continuous and not a single harmonic. 
We would like to point out that the main difficulty
with the theory of coronal loop heating by dissipation of slow
magnetoacoustic waves has been connected with the absence of the
observational evidence of these waves in the corona. It has been
believed that the waves somehow generated below the temperature
minimum, are not able to penetrate through this layer. Now, the observational
evidence of the long period longitudinal waves has been provided. The 
long period waves can have unresolved shorter period components in
their spectrum. According to the theory, these short period
spectral components dissipate very quickly, and deposit the energy
in the near footpoint part of the loop legs.

In this paper we develop further the model of Nakariakov et al. (2000) for
one-dimensional slow magnetoacoustic waves in stratified coronal loops,
taking into account the effects of the non-zero loop plane inclination angle and 
the semi-circular loop offset,
and derive a generalized evolutionary equation for slow magnetoacoustic waves
in coronal loops. It is shown that a coronal loop
acts as a dissipative filter on the slow magnetoacoustic waves and that
the short wavelength part of the wave spectrum dissipates in the
nearest vicinity of the loop footpoints. 
By calculating the heat deposition rate, provided by the dissipation of the
slow waves and comparing it with the observationally determined heat deposition
rate in EUV loops, we show that sufficiently wide spectrum slow waves 
can deposit the
required amount of energy  in the lower parts of the loops,
which is consistent with observations.

This paper is organized as follows: we formulated our generalized
model in Section~2; Section~3 deals with the comparison of our model's
predictions with the observations; and, finally, in Section 4 
we close with a discussion of our main results.

\section{The model}

\begin{figure}
     \resizebox{\hsize}{!}{\includegraphics{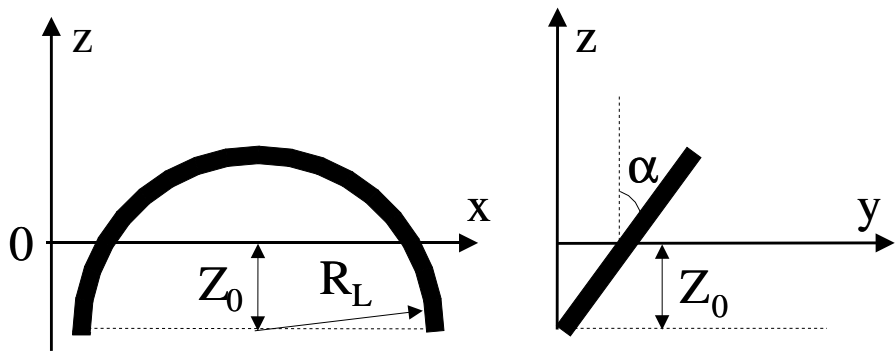}}
     \caption{The sketch of the model considered. A coronal loop is considered
     as a straight magnetic field line with density and gravitational 
     acceleration varying along
     the axis of the cylinder.}
\label{fig1}
\end{figure}

We consider a semi-circular loop of the curvature radius $R_L$.
The loop cross-section is taken to be constant. Validity of the
latter assumption has been corroborated by Klimchuk et al. (1992)
and by Aschwanden et al. (1999) for soft X-ray and EUV loops,
respectively. The loop is filled with a gravitationally stratified
magnetized plasma of constant temperature. Here, we generalize
Nakariakov et al.'s (2000) work by incorporating effects of loop
plane inclination and offset of circular loop centre from the
baseline. This is necessary since actually observed loops have the
non-zero (measured from the normal to the solar surface)
inclination angle $\alpha$ and non-zero offset of circular loop
centre from baseline, $Z_0$, i.e. distance from the circle's centre
to the solar surface. The sketch of the model geometry is shown
in Figure~1.

The gravitational acceleration projected on the loop tangential at
any point which is located at a distance $s$ measured from the
footpoint ($s=0$) along the loop, is
$$
g(s) = {{GM_{\sun}}\over {R^2_{\sun}}}\
 \left[\sqrt{1-{{Z_0^2}\over{R_L^2}}}
\cos\,\frac{s}{R_L}-{{Z_0}\over{R_L}}\sin\,\frac{s}{R_L} \right] \cos\,\alpha
$$
\begin{equation}
\times \left[1 + \cos\,\alpha
{{R_L}\over{R_{\sun}}} \left( \sqrt{1-{{Z_0^2}\over{R_L^2}}}
\sin\,\frac{s}{R_L}-{{Z_0}\over{R_L}} \left[ 1- \cos\,\frac{s}{R_L} \right]
\right) \right]^{-2},
\label{g}
\end{equation}
 where $G$ is the gravitational constant,
$R_L$ is the loop radius, $R_{\sun}$ and $M_{\sun}$ are solar
radius and mass, respectively.

\begin{figure}
     \resizebox{\hsize}{!}{\includegraphics{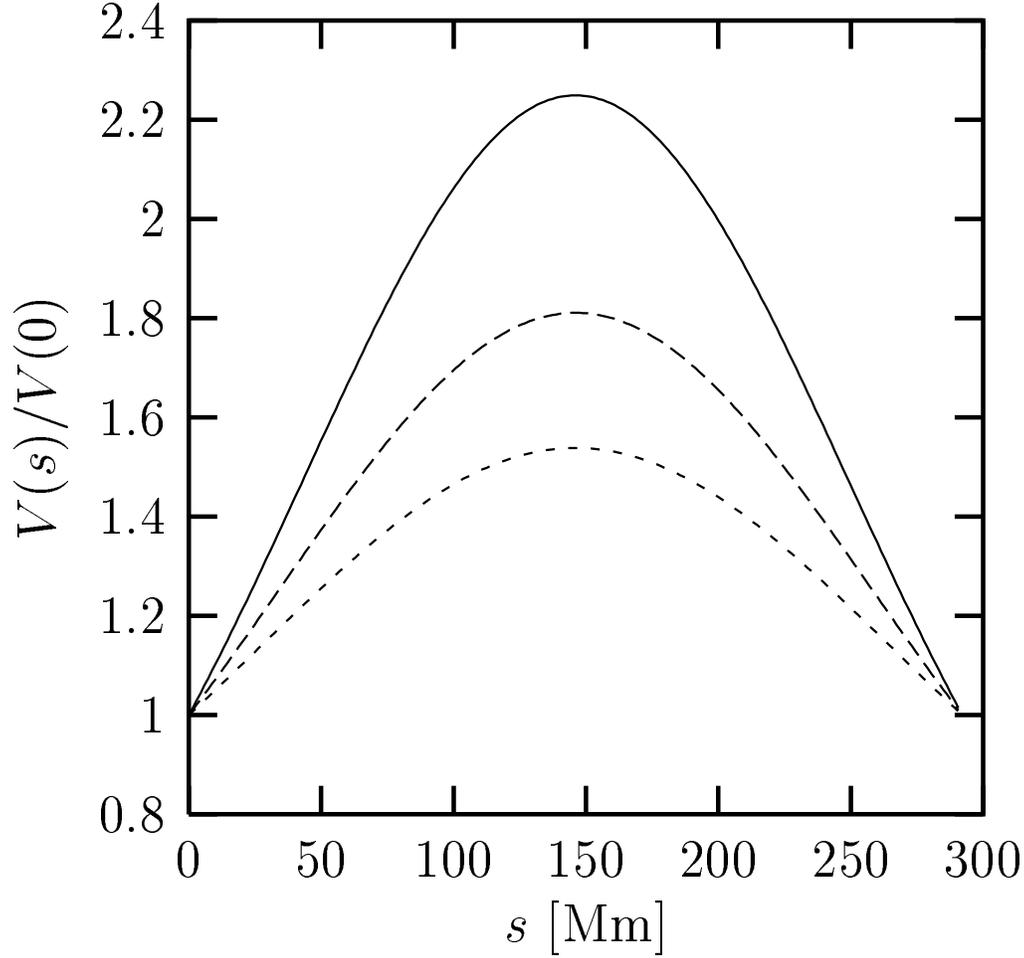}}
     \caption{Evolution of amplitude of a harmonic slow magnetoacoustic wave
     propagating along a coronal loop of the radius 93~Mm and zero off-set for
     different loop plane inclination angles.
     The wave period is 300~s. The medium is dissipationless, $\bar \eta =
     10^{-9}$. The solid curve corresponds to the inclination angle $\alpha=0$,
and long and short dashed lines represent $\alpha= \pi/ 4$ and
$\alpha = \pi/ 3$, respectively.}
\label{fig2}
\end{figure}

\begin{figure}
     \resizebox{\hsize}{!}{\includegraphics{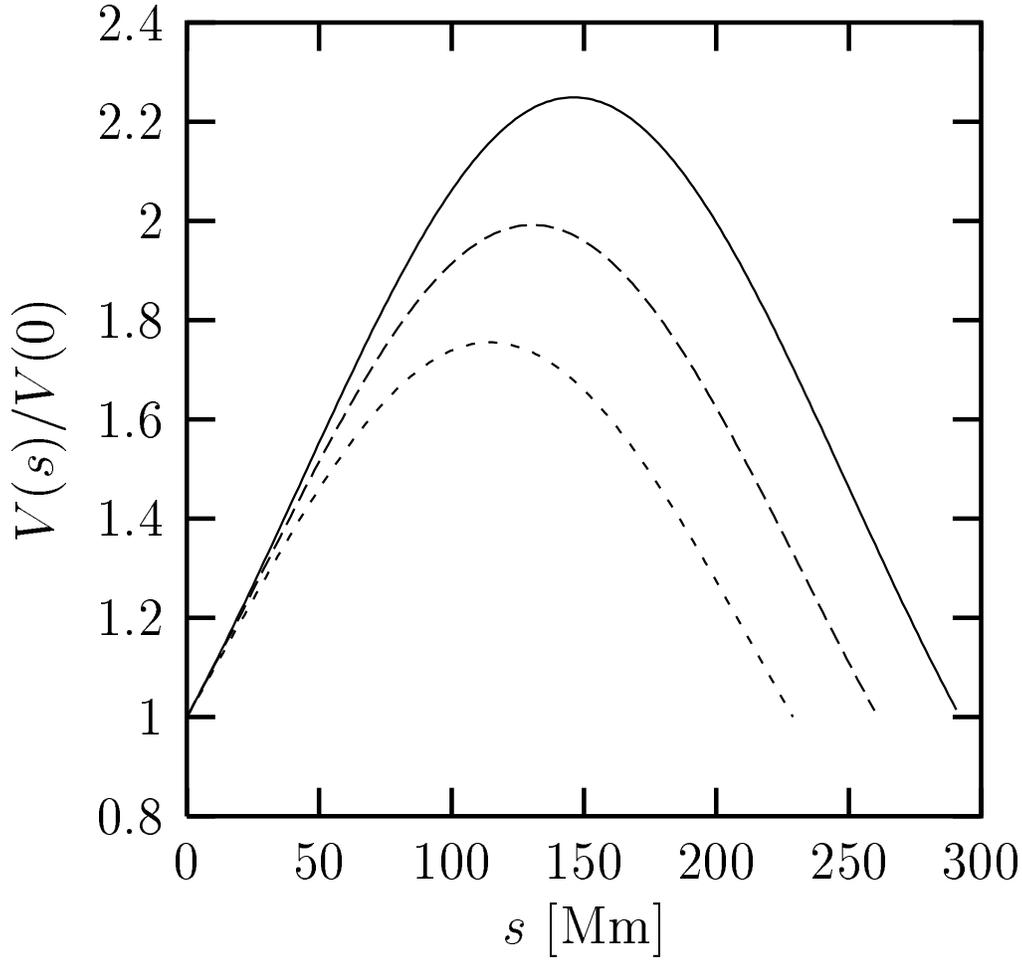}}
     \caption{Evolution of amplitude of a harmonic slow magnetoacoustic wave
     propagating along a coronal loop of the radius 93~Mm and zero inclination angle.
     The wave period is 300~s. The medium is dissipationless, $\bar \eta =
     10^{-9}$.
The solid curve corresponds to the loop with the off-set $Z_0=0$,
and long and short dashed lines represent $Z_0= R_L/ 4$ and $Z_0=
R_L/ 2$ respectively.} \label{fig3}
\end{figure}

The use of the hydrostatic equilibrium equations and the
isothermal equation of state, allows us to write the stationary
density profile along the loop as
\begin{equation}
\rho_0(s)=\rho_0(0)\exp \left( -{{\gamma}\over{C_s^2}} \int_0^s
g(s') ds'\right), \label{rho1}
\end{equation}
where $\gamma=5/3$ is the adiabatic index and $C_s$ is the speed
of sound. The temperature $T$ was assumed
to be 1~MK, giving the sound speed $C_s$ of 152~km/s.
An effective scale height can be determined as
$H(s) = C_s^2/\gamma g(s)$. 

Inserting  generalized
expression (\ref{g}) for $g(s)$ into (\ref{rho1}) we obtain
\begin{displaymath}
\rho_0(s)=\rho_0(0)\exp \Biggl[ -{{\gamma g(0)}\over{C_s^2}}
\end{displaymath}
\begin{equation}
\times {{\left[R_L
\sqrt{1-(Z_0/R_L)^2}\sin(s/R_L)-Z_0(1-\cos(s/R_L)) \right]
\cos\,\alpha} \over{1+\cos\,\alpha \left[(R_L /
{R_{\sun}})\sqrt{1-(Z_0 / R_L)^2} \sin(s/R_L)-(Z_0 /
{R_{\sun}})(1-\cos(s/R_L)) \right]} }\Biggl].
\label{rho2}
\end{equation}
Note, that when $\alpha \to
0$, expressions (\ref{g}) and (\ref{rho2})  reduce to the case
investigated by Nakariakov et al. (2000).

\begin{figure}
     \resizebox{\hsize}{!}{\includegraphics{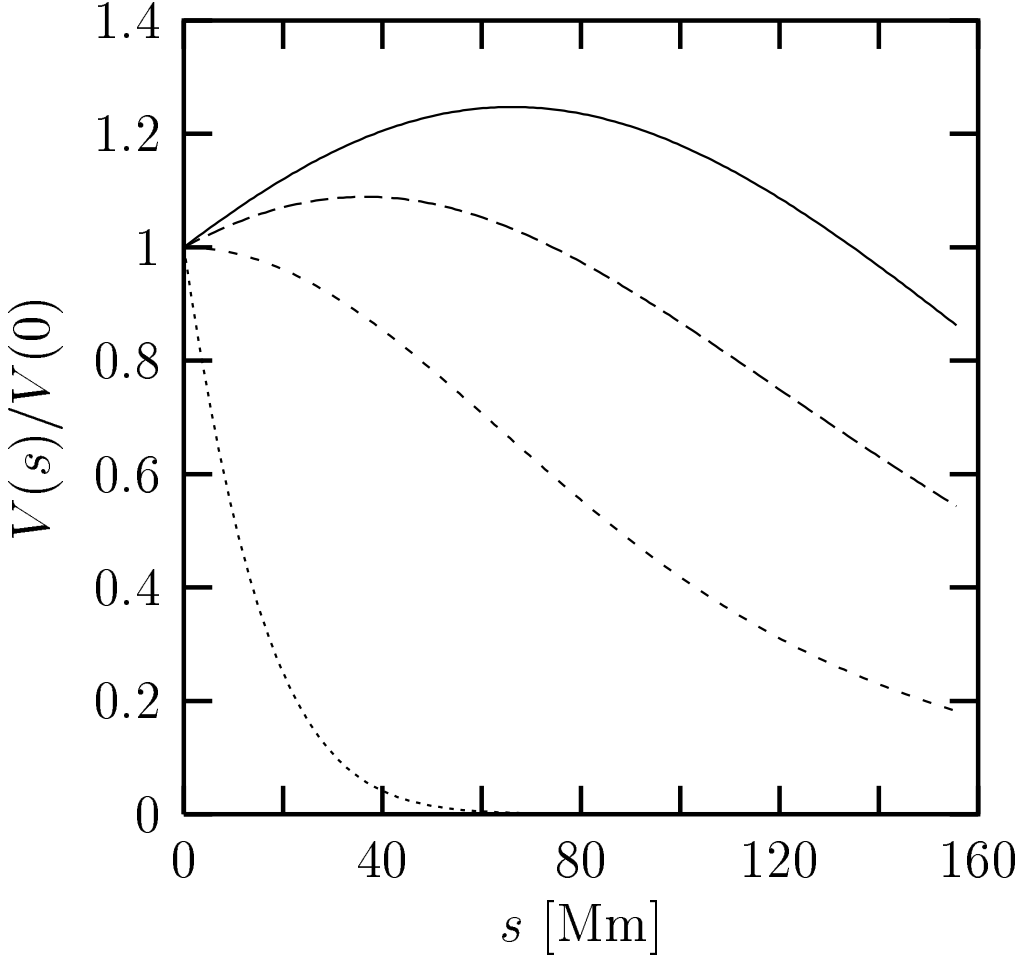}}
     \caption{Evolution of the slow wave amplitude in the average
     EUV loop (Aschwanden et al. 1999) with the parameters:
     $R_L= 93$~Mm, $Z_0=62$~Mm, $\alpha=7^{\degr}$. The normalized coefficient of
dissipation $\bar \eta$ is $4.0 \times 10^{-4}$. The solid curve
corresponds to the wave period of 600~s, while long dashed, 
short dashed and dotted curves correspond to 300~s, 120~s and 60~s
respectively.}
\label{fig4}
\end{figure}

For the following, we introduce dimensionless variables,
$\bar R_L=R_L/R_{\sun}$, $\bar Z_0=Z_0/R_{\sun}$, $\bar \rho
= \rho / \rho_0$, $\bar H = H/R_{\sun}$, $\bar V = V/C_s$ and $\bar s = s/R_{\sun}$.
Assuming that the wavelength is much less than both the scale height and
the dissipation length, and applying the multi-scale expansion method described
in detail in Nakariakov et al. (2000),
 we obtain the evolutionary equation for slow
magnetoacoustic waves,
\begin{equation}
{{\partial \bar V}\over{\partial \bar s}} -{{1}\over{2 \bar H(\bar s)}}
\bar V - {{\bar \eta}\over{2 \bar \rho(\bar s)}}{{\partial^2
\bar V}\over{\partial \xi^2}} =0,
\label{eveq}
\end{equation}
where $\xi = \bar s- C_s t / R_{\sun}$ is the running coordinate and
\begin{equation}
\bar \eta = \frac{1}{\rho_0(0) C_s R_\odot}\left[ \frac{4
\eta_0}{3} + \frac{\kappa_{||}(\gamma-1)^2}{{\cal R} \gamma}
\right],
\label{neta}
\end{equation}
is the normalized coefficient of dissipation connected with the
coefficient of thermal conduction $\kappa_{||}$ and the volume
viscosity $\eta_0$ (see the discussion in Nakariakov et al. 2000),
the rest of the notations are standard.

Eq.~(\ref{eveq}) together with (\ref{g}) and (\ref{rho2})
generalizes the equation derived by Nakariakov et al. (2000) to
the case of non-zero loop plane inclination and the offset, $Z_0$. But,
it does not contain the nonlinear term. This term is neglected
because it was shown that, for the amplitudes observed, the effect
of the nonlinearity is insignificant. Neglecting the
nonlinearity allows us to investigate solutions of the evolutionary
equation in all the details analytically.

Considering a harmonic wave, $\propto \exp(i k \xi)$, 
 we solve Eq.~(\ref{eveq}) for the amplitude of
the $k$-th spectral component,
\begin{equation}
 V(\bar s,k)=V(0,k)\exp \left[\phi_1(\bar s) -
k^2 \phi_2(\bar s) \right], 
\label{eveqs}
\end{equation}
where
\begin{displaymath}
\phi_1(\bar s)= \int_0^{\bar s}
{{1}\over{2\bar H(\bar s')}}
d \bar s'=
\end{displaymath}
\begin{equation}
\left({{\gamma g(0)}\over{2 C_s^2 R_{\sun}}} \right) {{\left[\bar
R_L \sqrt{1-(\bar Z_0/\bar R_L)^2}\sin(\bar s/\bar R_L)- \bar
Z_0(1-\cos(\bar s/ \bar R_L)) \right] \cos\,\alpha}
\over{1+\cos\,\alpha \left[\bar R_L \sqrt{1-(\bar Z_0 / \bar
R_L)^2} \sin(\bar s/ \bar R_L)-\bar Z_0 (1-\cos(\bar s/ \bar R_L))
\right]} }, \label{phi1}
\end{equation}
\begin{equation}
\phi_2(\bar s)= \int_0^{\bar s} {{\bar \eta}\over{2 \bar \rho(\bar
s')}} d \bar s', \label{phi2}
\end{equation}
and $k$ is the dimensionless wave number,
 $k = 2\pi R_{\sun} / \lambda = 2 \pi R_{\sun}/(C_s P)$ with
 $\lambda$ and $P$ being the wavelength and wave period respectively.
Here, we used the fact that short wavelength slow magnetoacoustic
waves are practically dispersionless and described by the
dispersion relation $\omega \approx C_s k$. Consequently, their
spatial and temporal spectra are similar.

The parametric studies of solution (\ref{eveqs}) are shown in
Figures~2 and 3. Fig.~2 demonstrates the behaviour of a harmonic
wave in identical loops with different values of the inclination
angle $\alpha$. Naturally, the increase of the inclination angle
decreases the value of the maximum amplitude the wave can reach.
Indeed, larger inclination angle means less stratification of the
plasma density in the loop, therefore lower maximal values. Fig.~3
shows the evolution of the wave amplitude along the loop for
different offsets of a circular loop centre from the baseline,
$Z_0$. The increase in the loop centre offset leads to the
decrease of the loop length and also decreases the angle between
the loop leg and the surface of the Sun at the footpoint. The
length of the loop is $L=2 R_L \arccos( Z_0 /R_L)$. Obviously, the
increase in $Z_0$ reduces the effective stratification of the loop
density and this results in lower maximal amplitudes reached by
the waves.

\section{Evolution of the spectrum}

Eq.~(\ref{eveqs}) describes evolution of the spectral component
$k$. In the exponent, the first term is positive in the ascending
stage of the wave propagation, becomes zero when the wave reaches
the loop apex and then becomes negative. The second term is
always negative. Consequently, in the ascending stage, the wave
evolution is described by the competition of these two terms
describing effects of stratification and dissipation,
respectively. Importantly, these two mechanisms operate in
different spectral domains. Mathematically, Eq.~(\ref{eveqs}) is
similar to an evolutionary equation of a dissipative filter in the
theory of radio circuits. Therefore, a coronal loop acts on slow
magnetoacoustic waves as a dissipative filter. If we knew the
initial spectrum of the wave, the equation would allow us to determine the
spectrum at each given height $s$ of the loop. Unfortunately, the
initial spectrum, especially its short wavelength part, cannot be
determined with the current resolution of EUV telescopes.

Evolution of the spectrum is accompanied by the dissipation of
spectral components. Different spectral components have different
effective scale heights of the dissipation. Assuming the initial
wave spectrum and following the spectrum evolution along the loop,
we can determine the energy deposition rate at each given height
$s$. Indeed, this function can be tested against the observations.

In order to compare predictions of our model with the
observational data, we have investigated the behaviour of different
slow magnetoacoustic wave 
spectral components in the {\it average EUV loop} determined by
Aschwanden et al. (1999), with $R_L= 93$~Mm, $Z_0=62$~Mm,
$\alpha=7^{\degr}$, corresponding to the loop length of about $L=156$~Mm. The
evolution of different spectral components is shown in Figure~4.
The dissipative coefficient $\bar \eta$ has the same value as was estimated
in (Nakariakov et al. 2000).
The spectral components corresponding to shorter wavelengths
(i.e. shorter period)
dissipate faster and closer to the loop footpoint.

The initial spectrum of the slow waves, $V(0,k) \equiv
\sqrt{f(k)}$ is unknown, but it should satisfy certain criteria. According to
observations, the spectrum should have a maximum in the vicinity
of 180--420~s. The velocity amplitude of the maximum is about
1--3\% of the speed of sound. Perhaps, there are also maxima in the
shorter period part of the spectrum, but they have not been
resolved observationally. So, we assume that the spectrum
gradually declines with the wave period. In order to avoid
singularities in the integration of the dissipated wave energy
over the whole spectrum $k = [0,+\infty)$, we assume that $f(k)
\to 0$ when $k \to + \infty$ and $k \to 0$. Thus, we choose the spectrum
$f(k)$ with a single maximum at $k_*$. A function which
satisfies these requirements is

\begin{equation}
f(k)=n^{n/2} \left(\frac{a}{2}\right)^{(n-1)/2} k\left({{n-1}\over{2}}a+
k^2\right)^{-n/2} \;\;\;\;\;\;\;\;\; (n \geq 3).
\label{spec}
\end{equation}
The spectra with $n=3,5,7$ are shown in Figure~5. For any $n$, the
spectrum $f(k)$ attains its maximal value at $k_*=\sqrt{a/2}$. For
$a=1.84\times 10^4$, 
$k_* \approx 96$, which corresponds to a wave period of 300~s, 
and $n=3$, the actual spectral peak ($f(k)> 0.69$) 
is situated between the wave numbers corresponding
to 1.5 and 12~min periods, which is consistent with observations. Note
that $f(k) \propto k^{-(n-1)}$ asymptotically as $k \to + \infty$.
The energy contained in the spectrum is determined by the spectral index
$n$. Higher values of the spectral index correspond to narrower spectra.

\begin{figure}
     \resizebox{\hsize}{!}{\includegraphics{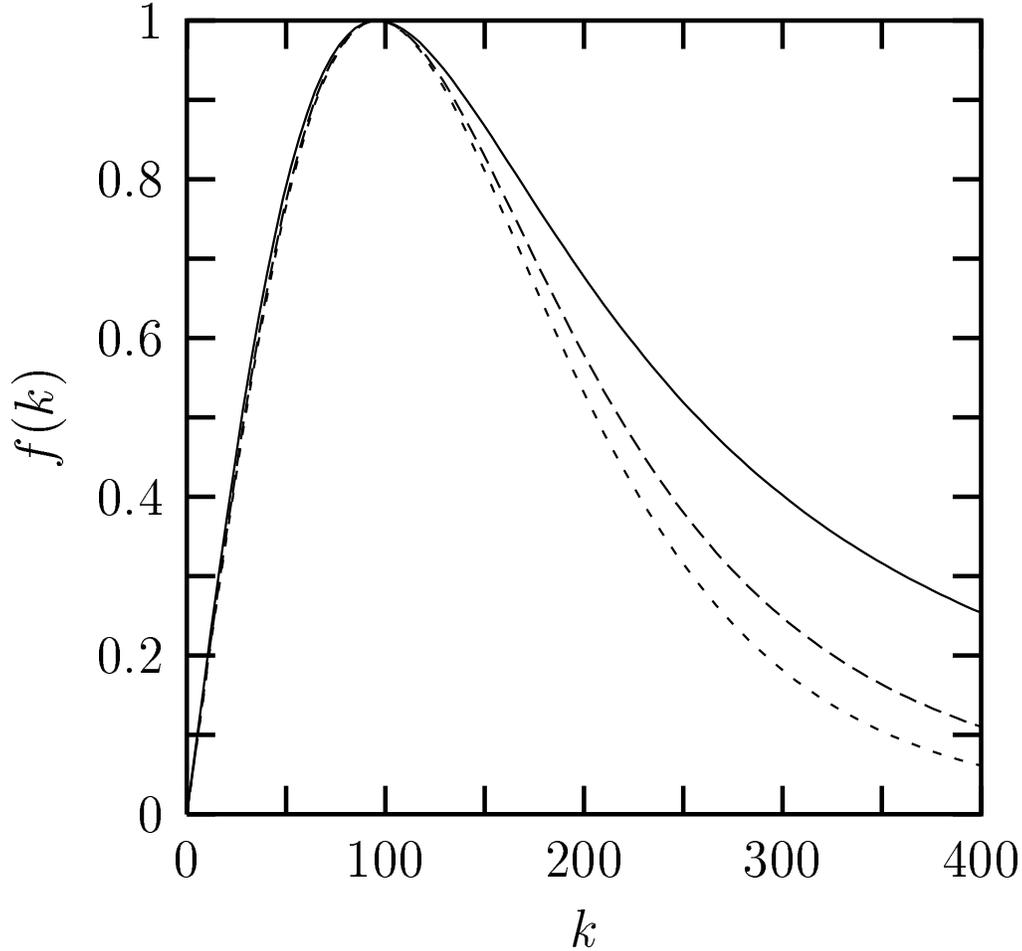}}
     \caption{Possible initial spectra $f(k)$ of slow magnetoacoustic waves.
     The solid curve corresponds to $n=3$, and long and short dashed
     curves to $n=5$ and 7, respectively. The dimensionless wave number $k$
     is connected with the wavelength $\lambda$ as $k = 2\pi R_{\sun}/\lambda$.}
     \label{fig5}
\end{figure}

The wave energy flux dissipated into heat in the loop segment from the
footpoint to the distance $s$ is defined as
\begin{equation}
F_D(s)=\int_0^{+ \infty} \left[F_{Dk}(0,k) - F_{Dk}(s,k) \right]\,
dk, 
\label{12}
\end{equation}
where
\begin{equation}
F_{Dk}(s,k)={{1}\over{2}}\rho_0(s)V^2(0,k) A^2 C_s^3 \exp \left[
2\phi_1(s)-2k^2\phi_2(s)\right]
\end{equation}
and $A$ is the amplitude of the strongest harmonic in the spectrum, 
measured in the units of the sound speed $C_s$, and Eq.~(\ref{eveqs}) was used.
To evaluate the integral in Eq.~(\ref{12}), we use the spectrum 
(\ref{spec}) and expressions (\ref{phi1}) and (\ref{phi2}). 
For the odd integer $n$, the integrals can be evaluated analytically, in particular
\begin{equation}
\int_0^{+ \infty} F_{Dk}^\mathrm{n=3}(s,k)\, dk = {{1}\over{2}} \rho_0 A^2C_s^3 (1.5
\sqrt{3}a)\biggl(a^{-1/2}- \sqrt{2 \pi \phi_2(s)}
e^{2\phi_2(s)a}\biggl[1- 
\label{in3}
\end{equation}
$$
\mathrm{erf}\left(\sqrt{2 a
\phi_2(s)}\right)\biggr] \biggr)
$$
\begin{equation}
\int_0^{+ \infty} F_{Dk}^\mathrm{n=5}(s,k)\, dk = {{1}\over{2}} \rho_0 A^2C_s^3
(5^{5/2}(a/2)^2/3)\biggl((2a)^{-3/2} +2  \sqrt{\pi} (2
\phi_2(s))^{3/2} e^{2\phi_2(s)(2a)} 
\label{in5}
\end{equation}
$$
\biggl[1- \mathrm{erf}\left(\sqrt{(2a) 2 \phi_2(s)}\right) \biggr] -2 (2
\phi_2(s))(2a)^{-1/2} \biggr)
$$
and
\begin{equation}
\int_0^{+ \infty} F_{Dk}^\mathrm{n=7}(s,k)\, dk= {{1}\over{2}} \rho_0 A^2C_s^3
(7^{7/2}(a/2)^3/15)\biggl(3(3a)^{-5/2} - 4  \sqrt{\pi}(2
\phi_2(s))^{5/2} e^{2\phi_2(s)(3a)} 
\label{in7}
\end{equation}
$$
\biggl[1-
\mathrm{erf}\left(\sqrt{ (3a)2
\phi_2(s)}\right)\biggr] -2 (2 \phi_2(s))(3a)^{-3/2} +
4(2 \phi_2(s))^2(3a)^{-1/2}\biggr)
$$
Here, $\mathrm{erf}(x)=(2/ \sqrt{\pi})\int_0^x \exp(-t^2) dt$ 
is the error function. 
In the evaluation of the integrals, we use that 
$\rho_0(s)\exp[2\phi_1(s)]=1$ and
\begin{equation}
I(n=3)=\int_0^{+ \infty} e^{-\mu x}/ \sqrt{(x+a)^3} \mathrm{d}x =
2/\sqrt{a} -2 \sqrt{\pi \mu} e^{a \mu} (1 -\mathrm{erf}(\sqrt{a
\mu})),
\label{14}
\end{equation}
\begin{equation}
I(n=5)=\int_0^{+ \infty} e^{-\mu x}/ \sqrt{(x+2a)^5} \mathrm{d}x =
{{2}\over{3}} \biggl(
(2a)^{-3/2} + 2 \sqrt{\pi} \mu^{3/2} e^{a \mu} (1 -
\label{15}
\end{equation}
$$
\mathrm{erf}(\sqrt{a
\mu}))-2 \mu (2a)^{-1/2}\biggr),
$$
\begin{equation}
I(n=7)=\int_0^{+ \infty} e^{-\mu x}/ \sqrt{(x+3a)^7} \mathrm{d}x =
{{2}\over{15}} \biggl(
3(3a)^{-5/2} - 4  \sqrt{\pi} \mu^{5/2} e^{a \mu} (1 -
\label{16}
\end{equation}
$$
\mathrm{erf}(\sqrt{a
\mu}))-2 \mu (3a)^{-3/2}+4 \mu^2(3a)^{-1/2}\biggr).
$$

The observationally determined rate of heat deposition at a
position $s$ of the loop can be approximated as
\begin{equation}
E_H(s)=E_{H0}\exp(-s/s_H), \label{obsEH}
\end{equation}
where  $E_{H0}$ is the heating rate at the footpoint and
$s_H$ is the heating scale height. The specific values 
of $E_{H0}$  and $s_H$ were determined by Aschwanden et al. (1999)
for EIT loops and by Aschwanden et al. (2000, 2001) for TRACE loops.
A physical value identical to $F_D(s)$
introduced in this paper, can be constructed by integration of
Eq.~(\ref{obsEH}) over $s$,
\begin{equation}
F_D^{obs}(s) \equiv \int_0^s E_H(s) ds = E_{H0}
 s_H (1-e^{-s/s_H}).
\label{FDobs}
\end{equation}
When $s=L/2$, at the loop apex, the value of the function $F_D^{obs}(s)$ 
is the total energy power introduced by Aschwanden et al. (2001, Eq.~(12)).

The comparison of the theoretical dependences of the wave energy flux 
dissipated into heat in a loop segment from the footpoint to the position $s$ given
by Eqs. (\ref{in3}), (\ref{in5}) and (\ref{in7}) and the observationally
determined dependence (\ref{FDobs}) is shown in Figure~6. 
Here, we use the parameters of nonuniform heating determined
in (Aschwanden et al. 2001), $E_{H0}=10^{-3.0\pm 0.3}$~erg
cm$^{-3}$~s$^{-1}$ and $s_H=12 \pm 5$~Mm. The normalized dissipation
is $\bar \eta =4.0 \times 10^4$. The density of plasma near the loop footpoint, 
which corresponds to the concentration at the base of the average loop, 
$n_0 = 1.92\times 10^9$~cm$^{-3}$, is $ \rho_0 = 4.1 \times 10^{-15}$ g/cm$^3$.
The amplitude $A$ of the wave, corresponding to the peak in the spectrum
(\ref{spec}), is taken to be $A=0.03$, i.e., 3\% of the sound speed.
The observational curve has the 
error bars determined by choosing plus and minus signs in the
$E_{H0}$ and $s_H$, given above.

\begin{figure}
     \resizebox{\hsize}{!}{\includegraphics{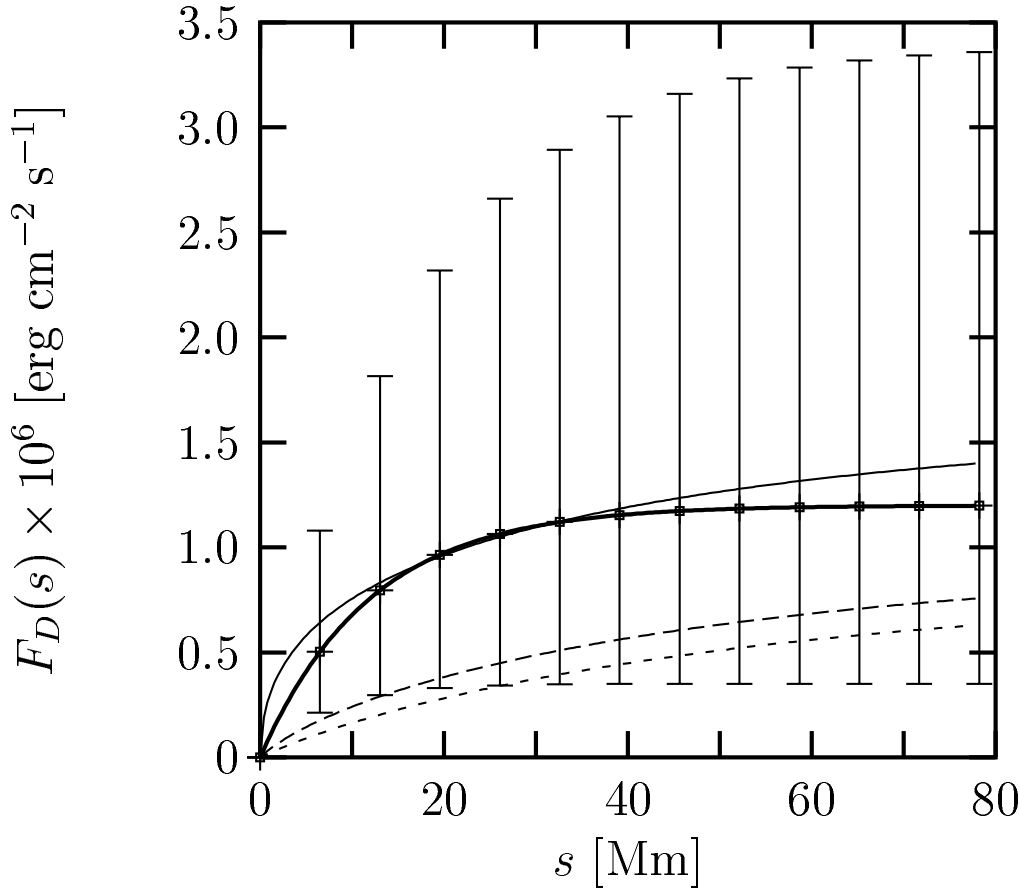}}
     \caption{The slow magnetoacoustic wave energy flux 
dissipated into heat in a loop segment from the footpoint to the position $s$
in comparison with the observationally determined nonuniform heating of an average loop.
The thin solid curve corresponds to the spectral index $n=3$, the long
dashed curve to $n=5$ and short dashed curve to $n=7$.
The observational dependence is the thick solid curve with the error bars.}
\label{fig6}
\end{figure}

Note that for $n=3$ the theoretical curve $F_D(s)$ fits 
the median value for the $F_{D}^{obs}(s)$ very well and 
that the curves corresponding to much steeper spectra, with
$n=5$ and $n=7$ are also almost inside the error range.

\section{Discussion}

Recently observed variations of EUV emission intensity, 
propagating upward along the 
legs of solar coronal loops, are satisfactory
interpreted in terms of the slow magnetoacoustic waves. We generalized
the model of the slow wave evolution in a loop, suggested by
Nakariakov et al. 2000 by taking into account effects of loop
plane inclination and offset of the semi-circular loop centre from
the solar surface. An evolutionary equation which governs the wave
propagation in presence of gravitational stratification and
dissipation was derived. Both non-zero inclination and 
offset affect the slow wave evolution, reducing the stratification of the
loop plasma and, consequently, the amplitude growth of the upward propagating waves.

It was found that coronal loops act as dissipative filters of slow
magnetoacoustic waves. Different spectral components have
different dissipation lengths and the wave spectrum evolves with
height or with the distance along the loop. Therefore, the
amount of wave energy deposited in the loop decreases with the
height and the wave dissipation deposits the energy in the lower
parts of the loops. Making certain assumptions about the 
spectrum of the waves near the loop footpoints, we constructed a
function describing the wave energy flux dissipated into heat in a 
loop segment. This function was compared with an observationally
determined function approximating the nonuniform heat deposition
requirements. We conclude,  
that the observed  heat deposition into coronal loops can be
interpreted as dissipation of slow magnetoacoustic waves, provided
the wave spectrum is sufficiently wide. This conclusion is consistent
with previously obtained results by Porter et al. (1994a,b), but
is based upon recently obtained observational information on the
actual existence and parameters of slow magnetoacoustic waves in coronal loops
and on the nonuniform distribution of heat deposition in EUV loops.  
In the scenario proposed, the short wave length components of the slow
wave spectrum, with short dissipation lengths,
are filtered out in the loop legs, depositing the
energy and heating the loops near the footpoints, while the long
wave length spectral components, with longer dissipation lengths,
propagate further and are detected by EUV telescopes.

Strictly speaking, realistic spectrum of slow magnetoacoustic
waves is restricted by the acoustic cutoff wave number from below
and by the inverse mean free path length from above. However, this
does not seem to affect the results obtained, as the impact of the
long wavelength (lower frequency) part of the spectrum to heating
is weak and as the short wavelength slow waves are efficiently
dissipated in the collisionless regime too, depositing the energy
near the loop footpoints.

Therefore, the theoretical results on the slow wave dissipation are 
consistent with the phenomenological results on EUV coronal loop
heating and with the observed amplitude of the waves in the long 
wavelength part of the spectrum. The wide spectrum slow 
magnetoacoustic waves can, indeed, deposit
sufficient amount of energy near the loop footpints, which is in 
agreement with the observations of Aschwanden et al. (1999, 2000, 2001). 
We would like to emphasize that
this possibility is purely theoretical 
due to not yet fully observed spectrum of the 
slow magnetoacoustic waves. However, our results indicate that
the slow waves cannot be excluded from the consideration 
as a possible mechanism for heating of the coronal loops.

\begin{acknowledgements} 
The authors would like to thank Markus Aschwanden and Erwin Verwichte
for valuable discussions.
\end{acknowledgements}

\end{document}